\documentclass[amsmath,amssymb,showpacs]{revtex4}

\usepackage{graphicx}
\usepackage{dcolumn}
\usepackage{bm}
\newcommand{\be}{\begin{equation}}
\newcommand{\ee}{\end{equation}}
\newcommand{\ba}{\begin{eqnarray}}
\newcommand{\ea}{\end{eqnarray}}
\newcommand{\baa}{\begin{eqnarray*}}
\newcommand{\eaa}{\end{eqnarray*}}

\begin{document}
   
\title{Replica-Exchange Molecular Dynamics Simulations for Various Constant Temperature Algorithms}
\author{%
Yoshiharu Mori$^{1}$ and 
Yuko Okamoto$^{1,2}$
}
\affiliation{$^{1}$Department of Physics, Nagoya University \\ Nagoya, Aichi 464-8602 \\
$^{2}$Structural Biology Research Center, Nagoya University \\ Nagoya, Aichi 464-8602}
\begin{abstract}
In the replica-exchange molecular dynamics method, where 
constant-temperature molecular dynamics simulations are performed in each replica, one usually rescales the momentum of each particle after replica exchange.
This rescaling method had previously been worked out only for the Gaussian constraint method.
In this letter, we present momentum rescaling formulae for four other commonly used constant-temperature algorithms, namely, Langevin dynamics, Andersen algorithm, Nos\'{e}-Hoover thermostat, and Nos\'{e}-Poincar\'{e} thermostat.
The effectiveness of these rescaling methods is tested with a small biomolecular system, 
and it is shown that proper momentum rescaling is necessary to obtain correct results 
in the canonical ensemble.
\end{abstract}

\maketitle

Monte Carlo (MC) and molecular dynamics (MD) simulations with generalized-ensemble algorithms have been widely used for studies of proteins and peptides (for a review, see, e.g., ref. 1).
Among generalized-ensemble algorithms, the replica-exchange method (REM)~\cite{hukushima96} (the method is also referred to as parallel tempering~\cite{marinari98}) is a useful simulation method because there is no need to determine the weight factor before the simulation.
The original REM was proposed for MC simulations and momenta of particles do not have to be considered,~\cite{hukushima96} but momenta have to be included in replica-exchange molecular dynamics (REMD) simulations and are usually rescaled when the exchange is performed.\cite{sugita99}
Although only dynamical variables of a physical system are considered and the usual Boltzmann weight factor is used in the detailed balance condition in the original REMD,\cite{sugita99} the states of the system can be specified by coordinates and momenta of atoms and additional dynamical variables, or the distribution function can be different from that of the canonical ensemble in some constant temperature algorithms.
For example, the Nos\'{e}-Hoover thermostat has a velocity of the thermostat, and the distribution function includes a kinetic energy term of this velocity.~\cite{hoover85}
Thus, one has to use a different rescaling method in REMD simulations depending on different constant temperature algorithms.

Let us consider a REM system, which consists of $M$ non-interacting replicas of the original system in the canonical ensemble at $M$ different temperature values $T_m \ (m=1, \cdots, M)$, and let $i \ (i=1, \cdots ,M)$ be a label which specifies the replica.
Then there is a one-to-one correspondence between the replica label and the temperature label and we can introduce a permutation function $\sigma$ and the inverse function $\sigma^{-1}$ as 
\begin{equation}
\begin{cases}
i = \sigma (m) , \quad m = \sigma^{-1}(i), \\
j = \sigma (n) , \quad n = \sigma^{-1}(j),
\end{cases}
\end{equation}
where $i$ and $j$ stand for the replica labels and $m$ and $n$ stand for the temperature labels, respectively.
The states of replica $i$ at temperature value $T_m$ are specified by $x^{[i]}_m \equiv (q^{[i]},p^{[i]},\alpha^{[i]})_m$.
Here, $q^{[i]} \equiv \{ \bm{q}_1^{[i]},\cdots ,\bm{q}_N^{[i]} \}$ and $p^{[i]}\equiv \{ \bm{p}_1^{[i]},\cdots ,\bm{p}_N^{[i]} \}$ are coordinates and momenta of $N$ atoms in replica $i$, respectively, and $\alpha^{[i]}$ is a set of additional dynamical variables depending on constant temperature algorithms in replica $i$.
The states of the REM system are specified by $X \equiv \{ x^{[\sigma (1)]}_1, \cdots , x^{[\sigma (M)]}_M \}$.

Suppose a pair of replicas $i$ and $j$ are exchanged and let $X'$ be a state after the replicas are exchanged.
Namely, we try to change the state of the REM system as follows:
\begin{equation}
	X = \{ \cdots , x^{[i]}_m, \cdots , x^{[j]}_n, \cdots \} \to X' = \{ \cdots , x^{[j]\prime}_m, \cdots , x^{[i]\prime}_n, \cdots \},
\end{equation}
where $x^{[j]\prime}_m$ and $x^{[i]\prime}_n$ are defined by
\begin{equation}
\begin{cases}
x^{[j]\prime}_m = \left( q^{[j]}, p^{[j]\prime}, \alpha ^{[j]\prime} \right) _m ,\\
x^{[i]\prime}_n = \left( q^{[i]}, p^{[i]\prime}, \alpha ^{[i]\prime} \right) _n ,
\end{cases}
\end{equation}
respectively, and explicit forms of the variables $p^{[i]\prime}, p^{[j]\prime}, \alpha ^{[i]\prime}$, and $\alpha ^{[j]\prime}$ are determined below.
After this transitions of states, the permutation function $\sigma$ is replaced by a new permutation function $\sigma '$ defined by
\begin{equation}
\begin{cases}
j = \sigma '(m) , \\
i = \sigma '(n) .
\end{cases}
\end{equation}

To make the REM system approach an equilibrium state, the detailed balance condition is imposed:
\begin{equation}
\label{dbc}
	P\left( X\right) w\left( X \to X'\right) = P\left( X'\right) w\left( X' \to X\right) ,
\end{equation}
where $w \left( X \to X'\right)$ is the transition probability from $X$ to $X'$ and
$P\left( X\right)$ is given by
\begin{equation}
	P\left( X\right) = \prod_{m=1}^M f_m(x^{[\sigma (m)]}_m) dx^{[\sigma (m)]}_m.
\end{equation}
Here, $f_m$ is a distribution function at temperature $T_m$.
The transition probability $w(X \to X')$ is then obtained by the Metropolis criterion:~\cite{metropolis53}
\begin{equation}
\label{transition_p}
	w\left( X \to X'\right) = \min \left[ 1, \frac{P(X')}{P(X)} \right] .
\end{equation} 

In general, the transition probability will be different from that of the replica-exchange MC algorithm, which is given by~\cite{hukushima96}
\begin{equation}
\label{remc1}
	w\left( X \to X'\right) = \min \left[ 1, \exp (-\varDelta ) \right] ,
\end{equation} 
where 
\begin{equation}
\label{remc2}
	\varDelta = (\beta_n - \beta_m )\left[ E(q^{[i]}) - E(q^{[j]}) \right] ,
\end{equation} 
$\beta_m$ is the inverse temperature defined by $\beta_m = 1 / k_{\text{B}}T_m$ ($k_{\text{B}}$ is the Boltzmann constant), and $E$ is the potential energy function.
However, we want this probabilty $w(X \to X')$ in eq. (\ref{transition_p}) of replica exchange to be independent of the constant-temperature algorithms.
One of such a choice is to use also the same probability as in eqs. (\ref{remc1}) and (\ref{remc2}).
We believe that this is the most natural choice.
Therefore, in the following discussion, we determine how to rescale the variables $p^{[i]\prime}, p^{[j]\prime}, \alpha ^{[i]\prime}$, and $\alpha ^{[j]\prime}$ for various constant temperature algorithms so that the transition probability is given by eqs. (\ref{remc1}) and (\ref{remc2}).
The constant temperature algorithms that are discussed in this letter are the Gaussian constraint method,~\cite{hoover82,evans83a,evans83b} Langevin dynamics,~\cite{allen87} Andersen algorithm,~\cite{andersen80} Nos\'{e}-Hoover thermostat,~\cite{hoover85} and Nos\'{e}-Poincar\'{e} thermostat.~\cite{bond99}

The Gaussian constraint method, in which the states are specified by $x \equiv (q,p)$, has the following distribution function:~\cite{evans83b}
\begin{equation}
	f(q,p) \propto \delta \left( \sum_{i=1}^N \frac{\bm{p}_i^2}{2m_i} - \frac{gk_{\text{B}}T}{2} \right) \exp \left[ -\beta E(q) \right] ,
\end{equation}
where $g = 3N-1$ if there are no constraints in the system.
Although the distribution function is not that of the canonical ensemble because the kinetic energy is fixed, it can be shown that the proper rescaling method in eqs. (\ref{remc1}) and (\ref{remc2}) is given by~\cite{sugita99}
\begin{equation}
\label{mom_scale}
	p^{[i]\prime} = \sqrt{\frac{T_n}{T_m}}p^{[i]}, \quad p^{[j]\prime} = \sqrt{\frac{T_m}{T_n}}p^{[j]}.
\end{equation}
This is the original momentum scaling introduced in the REMD method.~\cite{sugita99}

In the Langevin dynamics and the Andersen algorithm, the states are also specified by $x \equiv (q,p)$ and the distribution function is that of the canonical ensemble,~\cite{allen87,andersen80} that is
\begin{equation}
	f(q,p) \propto \exp \left[ -\beta \left( \sum_{i=1}^{N}\frac{\bm{p}_i^2}{2m_i} + E(q) \right) \right].
\end{equation}
The rescaling method in the algorithms is obtained, following the original REMD paper~\cite{sugita99} and again given by eq. (\ref{mom_scale}).

In the Nos\'{e}-Hoover thermostat, the states are specified by $x \equiv (q,p,\zeta )$ and the distribution function is given by~\cite{hoover85}
\begin{equation}
	f(q,p,\zeta) \propto \exp \left[ -\beta \left( \sum_{i=1}^N \frac{\bm{p}_i^2}{2m_i} + E(q) + \frac{Q}{2}\zeta^2 \right) \right],
\end{equation}
where $\zeta$ is a thermostat velocity and $Q$ is its mass parameter.
The mass parameter can have different values in each replica in REMD simulations.
It is found that the transition probability is given by eqs. (\ref{remc1}) and (\ref{remc2}), if the momenta are rescaled by eq. (\ref{mom_scale}) and $\zeta^{[i]\prime}$ and $\zeta^{[j]\prime}$ are rescaled by
\begin{equation}
	\zeta^{[i]\prime} = \sqrt{\frac{T_nQ_m}{T_mQ_n}}\zeta^{[i]}, \quad \zeta^{[j]\prime} = \sqrt{\frac{T_mQ_n}{T_nQ_m}}\zeta^{[j]},
\end{equation}
where $Q_m$ and $Q_n$ are the mass parameters in the replicas at temperature values $T_m$ and $T_n$, respectively. 
The rescaling method can be generalized to the Nos\'{e}-Hoover chains~\cite{martyna92} in a similar way.

In the Nos\'{e}-Poincar\'{e} thermostat, the states are specified by $x \equiv (q,\tilde{p},s,\pi)$ and the distribution function is given by \cite{bond99}
\begin{equation}
	f(q,\tilde{p},s,\pi )  \propto \delta \left[ s \left( H_{\text{N}} - \mathcal{E} \right) \right],
\end{equation}
where $\mathcal{E}$ is an initial value of $H_{\text{N}}$ and $H_{\text{N}}$ is the Nos\'{e} Hamiltonian,~\cite{nose84a,nose84b} which is given by
\begin{equation}
	H_{\text{N}} = \sum_{i=1}^N\frac{\tilde{\bm{p}}_i^2}{2m_is^2} + E(q) + \frac{\pi ^2}{2Q} + gk_{\text{B}}T\log s .
\end{equation}
Here, $g(=3N)$ is the number of degrees of freedom, $s$ is a position variable of the thermostat, $\pi$ is a momentum conjugate to $s$, and $\tilde{\bm{p}}_i$ is a virtual momentum, which is related to the real momenta $\bm{p}_i$ as $\bm{p}_i = \tilde{\bm{p}}_i / s$.
A rescaling method of the Nos\'{e}-Poincar\'{e} thermostat can be given by eq. (\ref{mom_scale}) and 
\begin{equation}
	\pi ^{[i]\prime} = \sqrt{\frac{T_nQ_n}{T_mQ_m}}\pi ^{[i]}, \quad \pi ^{[j]\prime} = \sqrt{\frac{T_mQ_m}{T_nQ_n}}\pi ^{[j]},
\end{equation}
\begin{equation}
	\label{np}
\begin{split}
	s^{[i]\prime} &= s^{[i]} \exp \left[ \frac{1}{gk_{\text{B}}} \left( \frac{E(q^{[i]})-\mathcal{E}_m}{T_m} - \frac{E(q^{[i]})-\mathcal{E}_n}{T_n} \right) \right] , \\
	s^{[j]\prime} &= s^{[j]} \exp \left[ \frac{1}{gk_{\text{B}}} \left( \frac{E(q^{[j]})-\mathcal{E}_n}{T_n} - \frac{E(q^{[j]})-\mathcal{E}_m}{T_m} \right) \right] ,
\end{split}
\end{equation}
where $\mathcal{E}_m$ and $\mathcal{E}_n$ are initial values of $H_\text{N}$ in the simulations with $T_m$ and $T_n$, respectively.
Note that the real momenta have to be used in the rescaling method in eq. (\ref{mom_scale}), not the virtual momenta.

In order to test the validity of the rescaling methods, we performed REMD simulations with the five constant temperature algorithms.
We used a system of a short peptide, Met-enkephalin, in gas phase, which has the amino-acid sequence Tyr-Gly-Gly-Phe-Met.
The N-terminus and the C-terminus of the peptide were blocked by the acetyl group and the N-methylamide group, respectively, and the initial structure was a fully extended one in all the replicas.
We used the following eight temperature values: $(T_1, \cdots , T_8) = (200,239,286,342,409,489,585,700)$ in kelvin.
Initial velocity of each atom was given by the Maxwell-Boltzmann distribution corresponding to the temperature in each replica.
The simulations were performed by the \textsc{tinker} program package.~\cite{ponder87}
Several of the programs in the package were modified and a few programs were added so that REMD simulations with the five constant temperature algorithms can be performed.
We used the AMBER parm99 force field~\cite{wang00}, and the dielectric constant was set to 1.0.
The unit time step was set to 0.5 fs and the simulation time was set to 5.0 ns for each replica.
We tried to exchange the replicas at every 10 fs and the data of the simulations were sampled just before the exchange trials.
The replica exchange was performed by exchanging the temperatures instead of the dynamical variables.
The pairs of replicas corresponding to neighboring temperature were simultaneously exchanged and the two choices of pairing, $(T_1,T_2),(T_3,T_4),\cdots$ and $(T_2,T_3),(T_4,T_5),\cdots$, were used alternately.
These conditions were common to all the simulations.
The computational details for each constant temperature algorithm are described below.

In the Gaussian constraint method, the integration method proposed by Zhang~\cite{zhang97} was used.
In the Langevin dynamics we used the integration method proposed by Mannella~\cite{mannella04} and the friction coefficient was set to 5.0 ps$^{-1}$.
In the Andersen algorithm the Velocity Verlet algorithm was used and the random collision frequency was set to 0.1 fs$^{-1}\cdot$atom$^{-1}$.
In the Nos\'{e}-Hoover thermostat we used the integration method proposed by Martyna \textit{et al.}~\cite{martyna96} and the values of mass parameters $Q$ were determined by the following equation~\cite{nose84b}:
\begin{equation}
\label{nosemass}
	Q = \frac{gk_{\text{B}}T}{\omega^2},
\end{equation}
where $\omega = 2\pi / \tau$ is a frequency of the thermostat and we set $\tau$ to 0.01 ps.
The initial values of $\zeta$ was set to 0.0.
In the Nos\'{e}-Poincar\'{e} thermostat we used the symplectic integrator proposed by Nos\'{e}~\cite{nose01}, which was recently generalized to biomolecular systems which include rigid-body molecules.~\cite{okumura07}
The value of each mass parameter was determined by the same way and had the same value as the Nos\'{e}-Hoover thermostat.
The initial values of $s$ and $\pi$ were set to 1.0 and 0.0, respectively.
For practical purposes, eq. (\ref{np}) cannot be used directly because the Nos\'{e} Hamiltonian is not actually conserved in the simulations.
Thus, modifications of eq. (\ref{np}) are required.
A simple solution is to replace $\mathcal{E}_m$ for replica $i$ and $\mathcal{E}_n$ for replica $j$ by the current value of the Nos\'{e} Hamiltonian.
This prescription works well as is shown below.

\begin{figure}[t]
\begin{center}
	\includegraphics[width=8cm,clip]{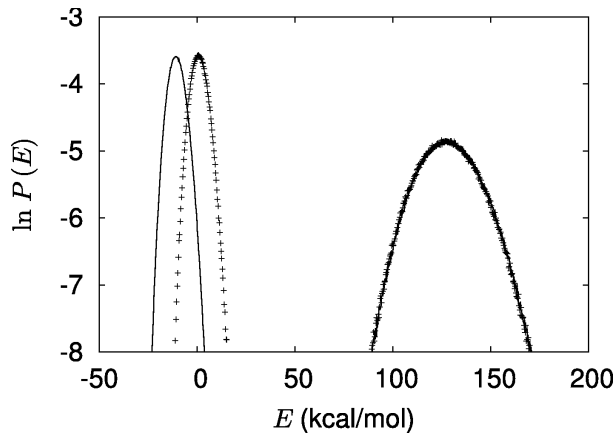}
\end{center}
\caption{Potential energy distributions in the simulations with and without REM for the Nos\'{e}-Poincar\'{e} thermostat.
The results with REM are represented by the solid curves and without REM by the crosses.
The distributions of the right side correspond to the highest temperature (700 K) and those of the left side correspond to the lowest temperature (200 K).}
\label{prob}
\end{figure}

\begin{figure}[b]
\begin{center}
	\includegraphics[width=8cm,clip]{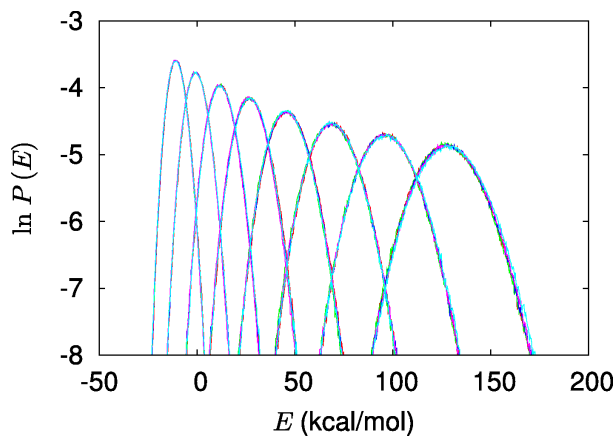}
\end{center}
\caption{(Color online) Potential energy distributions that were obtained by the REMD simulations with the five constant temperature algorithms. All of the results are plotted and essentially coincident among the five algorithms.}
\label{dist}
\end{figure}

\begin{figure}[b]
\begin{center}
	\includegraphics[width=8cm,clip]{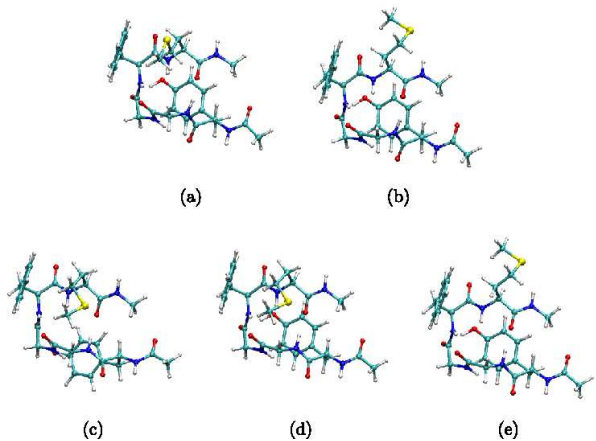}
\end{center}
\caption{(Color online) Structures of the lowest potential energy state that were obtained in the REMD simulations with (a) the Gaussian constraint method, (b) the Langevin dynamics, (c) the Andersen algorithm, (d) the Nos\'{e}-Hoover thermostat, and (e) the Nos\'{e}-Poincar\'{e} thermostat, respectively.
These figures were created by the \textsc{vmd} software.\cite{humphrey96}}
\label{struct}
\end{figure}

We also performed conventional canonical simulations (that is, without REM) with five constant temperature algorithms under the same conditions as in the REMD simulations so that the results with REM may be compared to those without REM.
Figure \ref{prob} shows the distributions of the potential energy in the simulations with and without REM for the Nos\'{e}-Poincar\'{e} thermostat.
At the highest temperature the distribution with REM agrees with that without REM, while at the lowest temperature the potential energy of the simulation with REM was lower than without REM.
Essentially the same results were also obtained for the other constant temperature algorithms and the obtained distributions for all the REMD simulations agreed with each other, which are shown in Fig. \ref{dist}.
The results at the lowest temperature by the canonical simulation without REM in Fig. \ref{prob} thus are wrong because it got trapped in states of local-minimum energy.
These results imply that REMD simulations with the rescaling methods can generate the correct canonical distributions.
Table \ref{ratio} lists the acceptance ratios for each REMD simulation.
The acceptance ratios are almost the same for all the simulations and therefore it is found that the efficiency of REMD simulations is independent of the kind of constant temperature algorithms.
\begin{table}
\caption{Acceptance ratios of replica exchange between pairs of the temperature values in the REMD simulations for all the constant temperature algorithms.
In this table, G, L, A, NH, and NP stand for Gauss, Langevin, Andersen, Nos\'{e}-Hoover, and Nos\'{e}-Poincar\'{e}, respectively.}
\label{ratio}
\begin{center}
	{
	\begin{tabular}{lccccccc} \hline \hline
		& $(T_1,T_2)$ & $(T_2,T_3)$ & $(T_3,T_4)$ & $(T_4,T_5)$ & $(T_5,T_6)$ & $(T_6,T_7)$ & $(T_7,T_8)$ \\ \hline
	    G & 0.148 & 0.140 & 0.137 & 0.132 & 0.129 & 0.131 & 0.139 \\
		L & 0.149 & 0.142 & 0.142 & 0.128 & 0.128 & 0.129 & 0.134 \\
		A & 0.151 & 0.143 & 0.141 & 0.130 & 0.125 & 0.128 & 0.135 \\
		NH & 0.151 & 0.144 & 0.139 & 0.130 & 0.127 & 0.132 & 0.138 \\
		NP & 0.151 & 0.144 & 0.142 & 0.132 & 0.127 & 0.128 & 0.136 \\ \hline \hline
	\end{tabular}}
\end{center}
\end{table}
Figure \ref{struct} shows the structures of the lowest potential energy state in all the REMD simulations.
Although the orientations of some side-chains are slightly different among the structures, the backbone structures were obtained with almost the same accuracy (less than 0.6 $\text{\AA}$ in rms deviation).

\begin{figure}[b]
\begin{center}
	\includegraphics[width=8cm,clip]{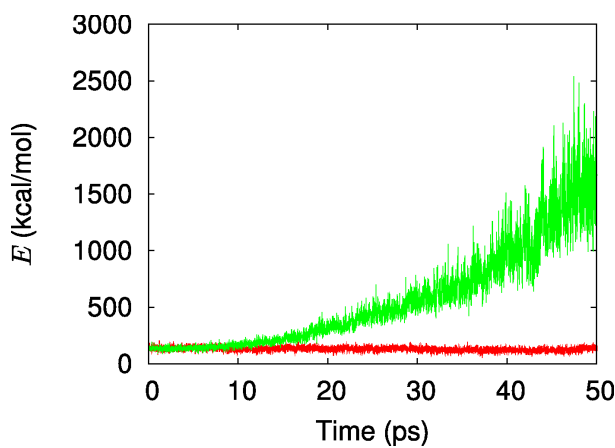}
\end{center}
\caption{(Color online) Time series of the potential energy in the REMD simulations at 700 K with the Nos\'{e}-Poincar\'{e} thermostat.
The results with the rescaling method are stable and flat (red curve), and those without the rescaling method are unstable (green curve).}
\label{noscale}
\end{figure}

To show that the rescaling methods are necessary to perform proper REMD simulations, we performed a REMD simulation with the Nos\'{e}-Poincar\'{e} thermostat, in which eqs. (\ref{remc1}) and (\ref{remc2}) were used but the dynamical variable $s$ was not rescaled.   
Figure \ref{noscale} shows the time series of the potential energy in the simulations when the dynamical variable $s$ was rescaled and not rescaled.
If $s$ was rescaled properly, the REMD simulation worked properly but otherwise the potential energy tended to diverge.
From these results, it is found that if dynamical variables are not properly rescaled in REMD simulations, then the simulation sometimes cannot provide correct results.
Therefore, one has to use the appropriate rescaling methods for each constant temperature algorithm when REMD simulations are performed.

In this letter, we proposed the rescaling methods which treat dynamical variables properly in REMD simulations for various constant temperature algorithms.
REMD simulations with the proper rescaling methods can provide correct canonical-ensemble distributions.  
With these rescaling methods for familiar constant temperature algorithms, REMD methods will become more applicable for simulations of various molecular systems.

\section*{Acknowledgments}

Some of the computations were performed on the supercomputers at the Information Technology Center, Nagoya University and at the Research Center for Computational Science, Institute for Molecular Science.
This work was supported, in part, by Grants-in-Aid for Scientific Research on Innovative Areas (``Fluctuations and Biological Functions") and for the Next Generation Super Computing Project, Nanoscience Program from the Ministry of Education, Culture, Sports, Science and Technology (MEXT), Japan.


\begin{thebibliography}{99}
	\bibitem{mitsutake01} A. Mitsutake, Y. Sugita, and Y. Okamoto: Biopolymers (Peptide Science) \textbf{60} (2001) 96.
	\bibitem{hukushima96} K. Hukushima and K. Nemoto: J. Phys. Soc. Jpn. \textbf{65} (1996) 1604.
	\bibitem{marinari98} E. Marinari, G. Parisi, and J. J. Ruiz-Lorenzo: in \textit{Spin Glasses and Random Fields}, ed. A. P. Young (World Scientific, Singapore, 1998) p. 59.
	\bibitem{sugita99} Y. Sugita and Y. Okamoto: Chem. Phys. Lett. \textbf{314} (1999) 141.
	\bibitem{hoover85} W. G. Hoover: Phys. Rev. A \textbf{31} (1985) 1695.
	\bibitem{metropolis53} N. Metropolis, A. W. Rosenbluth, M. N. Rosenbluth, A. H. Teller, and E. Teller: J. Chem. Phys. \textbf{21} (1953) 1087.
	\bibitem{hoover82} W. G. Hoover, A. J. C. Ladd, and B. Moran: Phys. Rev. Lett. \textbf{48} (1982) 1818.
	\bibitem{evans83a} D. J. Evans: J. Chem. Phys. \textbf{78} (1983) 3297.
	\bibitem{evans83b} D. J. Evans and G. P. Morriss: Phys. Lett. \textbf{98A} (1983) 433.
	\bibitem{allen87} M. P. Allen and D. J. Tildesley: \textit{Computer Simulation of Liquids} (Oxford, New York, 1987) p. 259. 
	\bibitem{andersen80} H. C. Andersen: J. Chem. Phys. \textbf{72} (1980) 2384.
	\bibitem{bond99} S. D. Bond, B. J. Leimkuhler, and B. B. Laird: J. Comput. Phys. \textbf{151} (1999) 114.
	\bibitem{martyna92} G. J. Martyna, M. L. Klein, and M. Tuckerman: J. Chem. Phys. \textbf{97} (1992) 2635.
	\bibitem{nose84a} S. Nos\'{e}: Mol. Phys. \textbf{52} (1984) 255.
	\bibitem{nose84b} S. Nos\'{e}: J. Chem. Phys. \textbf{81} (1984) 511.
	\bibitem{ponder87} J. W. Ponder and F. M. Richards: J. Comput. Chem. \textbf{8} (1987) 1016.
	\bibitem{wang00} J. Wang, P. Cieplak, and P. A. Kollman: J. Comput. Chem. \textbf{21} (2000) 1049.
	\bibitem{zhang97} F. Zhang: J. Chem. Phys. \textbf{106} (1997) 6102.
	\bibitem{mannella04} R. Mannella: Phys. Rev. E \textbf{69} (2004) 041107.
	\bibitem{martyna96} G. J. Martyna, M. E. Tuckerman, D. J. Tobias, and M. L. Klein: Mol. Phys. \textbf{87} (1996) 1117.
	\bibitem{nose01} S. Nos\'{e}: J. Phys. Soc. Jpn. \textbf{70} (2001) 75.
	\bibitem{okumura07} H. Okumura, S. G. Itoh, and Y. Okamoto: J. Chem. Phys. \textbf{126} (2007) 084103.
	\bibitem{humphrey96} W. Humphrey, A. Dalke, and K. Schulten: J. Molec. Graphics \textbf{14} (1996) 33.
\end{thebibliography}
\end{document}